\documentclass[12pt,a4paper]{article}

\def\hybrid{\topmargin -20pt    \oddsidemargin 0pt
        \headheight 0pt \headsep 0pt
        \textwidth 6.25in       
        \textheight 9.5in       
        \marginparwidth .875in
        \parskip 5pt plus 1pt   \jot = 1.5ex}

\hybrid

\def\baselinestretch{1.2}

\catcode`\@=11

\def\marginnote#1{}
\newcount\hour
\newcount\minute
\newtoks\amorpm
\hour=\time\divide\hour by60
\minute=\time{\multiply\hour by60 \global\advance\minute by-\hour}
\edef\standardtime{{\ifnum\hour<12 \global\amorpm={am}%
        \else\global\amorpm={pm}\advance\hour by-12 \fi
        \ifnum\hour=0 \hour=12 \fi
        \number\hour:\ifnum\minute<10 0\fi\number\minute\the\amorpm}}
\edef\militarytime{\number\hour:\ifnum\minute<10 0\fi\number\minute}

\def\draftlabel#1{{\@bsphack\if@filesw {\let\thepage\relax
   \xdef\@gtempa{\write\@auxout{\string
      \newlabel{#1}{{\@currentlabel}{\thepage}}}}}\@gtempa
   \if@nobreak \ifvmode\nobreak\fi\fi\fi\@esphack}
        \gdef\@eqnlabel{#1}}
\def\@eqnlabel{}
\def\@vacuum{}
\def\draftmarginnote#1{\marginpar{\raggedright\scriptsize\tt#1}}

\def\draft{\oddsidemargin -.5truein
        \def\@oddfoot{\sl preliminary draft \hfil
        \rm\thepage\hfil\sl\today\quad\militarytime}
        \let\@evenfoot\@oddfoot \overfullrule 3pt
        \let\label=\draftlabel
        \let\marginnote=\draftmarginnote
   \def\@eqnnum{(\theequation)\rlap{\kern\marginparsep\tt\@eqnlabel}%
\global\let\@eqnlabel\@vacuum}  }

\def\preprint{\twocolumn\sloppy\flushbottom\parindent 2em
        \leftmargini 2em\leftmarginv .5em\leftmarginvi .5em
        \oddsidemargin -.5in    \evensidemargin -.5in
        \columnsep .4in \footheight 0pt
        \textwidth 10.in        \topmargin  -.4in
        \headheight 12pt \topskip .4in
        \textheight 6.9in \footskip 0pt
        \def\@oddhead{\thepage\hfil\addtocounter{page}{1}\thepage}
        \let\@evenhead\@oddhead \def\@oddfoot{} \def\@evenfoot{} }

\def\numberbysection{\@addtoreset{equation}{section}
        \def\theequation{\thesection.\arabic{equation}}}

\def\underline#1{\relax\ifmmode\@@underline#1\else
        $\@@underline{\hbox{#1}}$\relax\fi}

\def\titlepage{\@restonecolfalse\if@twocolumn\@restonecoltrue\onecolumn
     \else \newpage \fi \thispagestyle{empty}\c@page\z@
        \def\thefootnote{\fnsymbol{footnote}} }

\def\endtitlepage{\if@restonecol\twocolumn \else \newpage \fi
        \def\thefootnote{\arabic{footnote}}
        \setcounter{footnote}{0}}  

\catcode`@=12
\relax

\def\figcap{\section*{Figure Captions\markboth
        {FIGURECAPTIONS}{FIGURECAPTIONS}}\list
        {Figure \arabic{enumi}:\hfill}{\settowidth\labelwidth{Figure
999:}
        \leftmargin\labelwidth
        \advance\leftmargin\labelsep\usecounter{enumi}}}
 \relax
\def\tablecap{\section*{Table Captions\markboth
        {TABLECAPTIONS}{TABLECAPTIONS}}\list
        {Table \arabic{enumi}:\hfill}{\settowidth\labelwidth{Table
999:}
        \leftmargin\labelwidth
        \advance\leftmargin\labelsep\usecounter{enumi}}}
 \relax
\def\reflist{\section*{References\markboth
        {REFLIST}{REFLIST}}\list
        {[\arabic{enumi}]\hfill}{\settowidth\labelwidth{[999]}
        \leftmargin\labelwidth
        \advance\leftmargin\labelsep\usecounter{enumi}}}
 \relax
\makeatletter
\newcounter{pubctr}
\def\publist{\@ifnextchar[{\@publist}{\@@publist}}
\def\@publist[#1]{\list
        {[\arabic{pubctr}]\hfill}{\settowidth\labelwidth{[999]}
        \leftmargin\labelwidth
        \advance\leftmargin\labelsep
        \@nmbrlisttrue\def\@listctr{pubctr}
        \setcounter{pubctr}{#1}\addtocounter{pubctr}{-1}}}
\def\@@publist{\list
        {[\arabic{pubctr}]\hfill}{\settowidth\labelwidth{[999]}
        \leftmargin\labelwidth
        \advance\leftmargin\labelsep
        \@nmbrlisttrue\def\@listctr{pubctr}}}
 \relax
\makeatother
\newskip\humongous \humongous=0pt plus 1000pt minus 1000pt

\newif\ifdtup

\relax

\def\be{\begin{equation}}
\def\ee{\end{equation}}
\def\ba{\begin{eqnarray}}
\def\ea{\end{eqnarray}}

\def\del{\partial}

\def\r{\rho}
\def\a{\alpha}

\def\d{\delta}

\def\e{\epsilon}

\def\th{\theta}

\def\m{\mu}

\def\om{\omega}
\def\Om{\Omega}

\def\L{\Lambda}
 
\def\S{\Sigma}
\def\vphi{\varphi}

\def\no{\noindent}

\def\qq{\qquad}
\def\bl{\bigl}
\def\br{\bigr}

\def\IR{\relax{\rm I\kern-.18em R}}

\def \ha {{1\over 2}}

\def \ov {\over}

\def\IR{\relax{\rm I\kern-.18em R}}
\def\inv{^{\raise.15ex\hbox{${\scriptscriptstyle -}$}\kern-.05em 1}}

\renewcommand{\theequation}{\arabic{equation}}

\begin{document}
\renewcommand{\theequation}{\arabic{equation}}

\newcommand{\beq}{\begin{equation}}
\newcommand{\eeq}[1]{\label{#1}\end{equation}}
\newcommand{\ber}{\begin{eqnarray}}
\newcommand{\eer}[1]{\label{#1}\end{eqnarray}}
\newcommand{\eqn}[1]{(\ref{#1})}
\begin{titlepage}
\begin{center}

\hfill CERN-TH/98-408\\
\hfill hep--th/9812165\\

\vskip .8in

{\large \bf {On (multi-)center branes and exact string vacua}\footnote{
To be published in the proceedings of the 
{\it Quantum Aspects of Gauge Theories,
Supersymmetry and Unification}, Corfu, Greece, 20--26 September 1998.}}

\vskip 0.4in

{\bf Konstadinos Sfetsos}
\vskip 0.1in
{\em Theory Division, CERN\\
     CH-1211 Geneva 23, Switzerland\\
{\tt sfetsos@mail.cern.ch}}\\
\vskip .2in

\end{center}

\vskip .4in

\centerline{\bf Abstract }

\vskip 0,2cm
\no
Multicenter supergravity solutions corresponding to Higgs phases 
of supersymmetric Yang--Mills theories are considered.
For NS5 branes we identify three cases where there is a  
description in terms of exact conformal field theories.
Other supergravity solutions, such as D3-branes with angular momentum,
are understood as special limits of multicenter ones.
Within our context we also consider 4-dim gravitational multi-instantons.

\vskip 7.5cm
\noindent
CERN-TH/98-408\\
December 1998\\
\end{titlepage}
\vfill
\eject

\def\baselinestretch{1.2}
\baselineskip 16 pt
\noindent

\section{Introduction}

It has been conjectured that strong coupling issues
in supersymmetric Yang--Mills (SYM) theories 
can be investigated using string theory on backgrounds containing AdS spaces 
\cite{Maldacena} (see also \cite{prrr}).
In this note, based to an extend on \cite{sfehi},
we present some supergravity solutions that 
describe the strong coupling 
regime of (SYM) theories for $SU(N)$
in the Higgs phase, where the gauge group is broken.
We show that in the case of NS5-branes, where R--R fields are absent, there 
exist exact conformal field theory (CFT) descriptions in a number of 
interesting cases.
The well known example corresponds to the $SU(2)\times U(1)$ WZW model 
CFT \cite{sl2u1}, representing the throat of a 
semi-wormhole, as the near horizon geometry of NS5 branes  
\cite{cahastro} at one center.
A less known example corresponds to a specific limit
of non-extremal NS5 branes (at one center), where the string 
coupling and non-extremality parameter both go to zero with their ratios 
held fixed. In this case the exact 
CFT description is in terms of $SL(2,\IR)/SO(1,1) \times SU(2)$ \cite{nonens}.
The first factor corresponds to the famous 2-dim black hole solution
\cite{witbla}. A third example \cite{sfehi} is that of the 
coset CFT $SU(2)/U(1) \times SL(2,\IR)/U(1)$ describing
branes uniformly distributed on a circle whose radius dictates the Higgs
expectation value for the scalars in the SYM theory side.
We also present a solution that interpolates between the first and third cases.
In addition, we give a new interpretation for the BPS limits of 
rotating D3-brane solutions as special limits of multicenter ones.
We have also written an appendix discussing 
(multi)-Eguchi--Hanson and (multi)-Taub--NUT metrics with removable NUT
singularities at points uniformly distributed on a circle.

\section{Branes on a circle}


Consider a
$d$-dim supergravity solution corresponding to $Nk$ parallel 
$p$-branes, which are separated into $N$ groups, 
with $k$ branes each, and have centers 
at $\vec x=\vec x_i$, $i=1,2,\cdots ,N$. It is
characterized by a harmonic function 
with respect to the $(n+2)$-dim space $E^{n+2}$,
which is transverse to the branes 
\be
H_n= 1 + \sum_{i=1}^N {a k C(\vec x_i)
\ov |\vec x - \vec x_i|^n}\ ,\qq n=d-p-3 \ ,\qq \sum_{i=1}^n C(\vec x_i) = N\ ,
\label{haH}
\ee
where $a$ is a constant that may depend,
according to the type of brane we discuss, only on the Planck length 
$l_{\rm P}$, the string 
length $l_s$ and the dimensionless string coupling constant $g_s$.
For a generic choice of vectors $\vec x_i$ and weight-function $C(\vec x_i)$, 
the $SO(n+2)$ symmetry of the
transverse space is broken.
Here we will make the simple choice that all the centers lie in 
a ring of radius $r_0$ in the plane defined by $x_{n+1}$ and $x_{n+2}$
and also that $\vec x_i = (0,0,\cdots , r_0 \cos\phi_i, r_0 \sin\phi_i)$, 
with $\phi_i = 2\pi i/N$. Moreover we choose $C(\vec x_i)=1$.
Hence the $SO(n+2)$ 
symmetry of the transverse space is broken to $SO(n) \times Z_N$.
Since the $\vec x_i$'s correspond to non-zero vacuum expectation values (vev's)
for the scalars, the
corresponding super Yang--Mills theory is broken from 
$SU(kN )$ to $U(k)^N$, with the vacuum having a $Z_N$ symmetry.
Then (\ref{haH}) can be written as 
\ba
&& H_n = 1 + a k \sum_{i=0}^{N-1} \bl(r^2 + r_0^2 
- 2 r_0 \r \cos(2\pi i/N - \psi)\br)^{-n/2}\ ,
\nonumber \\
&& r^2= \vec x^2\ ,\qq x_{n+1} = \r \cos\psi,\quad x_{n+2}= \r \sin\psi \ .
\label{summm}
\ea
The finite sum in (\ref{summm}) can be computed for even (odd) $n$
from that for $n=2$ ($n=1$).
In this way we find the harmonic corresponding to $N$ D5 (NS5) branes
on a circle of radius $r_0$ to be \cite{sfehi}
\be 
H_2 = 1 + {kN l_s^2 g_s^\d \ov 
\bl((r^2+r_0^2)^2-4 r_0^2 \r^2\br)^{1/2}}\ \L_N(x,\psi)\ , 
\label{hans55}
\ee
where 
\ba
 \L_N& \equiv& {\sinh(Nx) \ov \cosh(Nx) - \cos(N\psi)}
\ = \ha\left(\coth\bl(N(x+i \psi)\br) + \coth\bl(N (x-i\psi)\br)\right) 
\nonumber \\
& = &  1+ \sum_{m\neq 0} e^{-N (|m| x-i m \psi)}\ ,
\label{lnxpsi}
\ea
and $\d=1\ (0)$ for D5 (NS5) branes. Similarly, the 
harmonic corresponding to $N$ D3 branes on a circle of radius $r_0$ 
is \cite{sfehi}
\be
 H_4 = 1 + 
{4\pi N k g_s l_s^4 (r^2+r_0^2)\ov \bl((r^2+r_0^2)^2-4 r_0^2 \r^2\br)^{3/2}}
\ \S_N(x,\psi) \ ,
\label{h44}
\ee
where
\ba
\S_N & \equiv &
\L_N +N {\left( (r^2+r_0^2)^2 - 4 r_0^2 \r^2\right)^{1/2}\ov r^2+r_0^2}\ 
{\cosh N x \cos N\psi -1 \ov (\cosh Nx -\cos N\psi )^2} 
\nonumber \\
& = & 
 1+ \sum_{m\neq 0} \left(1+ {\bl((r^2+r_0^2)^2-4 r_0^2\r^2\br)^{1/2}
\ov r^2 + r_0^2}\ N |m| \right) e^{-N(|m| x - i m \psi)}\ .
\label{jhlk}
\ea
The variable $x$ appearing in (\ref{lnxpsi}), (\ref{jhlk}) is defined as 
\be
e^x \equiv {r^2 + r_0^2 \ov 2 r_0 \r} 
+ \sqrt{\left(r^2 + r_0^2 \ov 2 r_0 \r\right)^2-1}\ .
\label{exx}
\ee
Note that in both (\ref{hans55}) and (\ref{h44}) there is an 
explicit $Z_N$ invariance under shifts of $\psi\to \psi + {2 \pi\ov N}$. 
In the $1\ov N$-expansion, $\L_N$ and $\Sigma_N$ have only a ``tree-level'' 
contribution, whereas the rest of the terms in the infinite sum 
are non-perturbative. 
In particular, the exponential factors $N (|m|x - m\psi)$ are 
likely to originate from 
configurations of the spontaneously broken
gauge theory that interpolate between the $N$ different degenerate vacua. 
It would be interesting, in the case of D3 branes, to find 
an interpretation in terms of configurations
in the ${\cal N}=4$ spontaneously broken SYM theory.
For odd $n$'s we have not been able to perform the corresponding 
finite sums exactly.
However, these can be computed in the large $N$ limit. Then 
we may replace the 
sum in (\ref{summm}) by an integral and give the result, for any $n$, 
in terms of a hypergeometric function 
\ba
H_n & \approx & 1+a k N \int_0^{2\pi} {d\phi\ov 2 \pi} (r^2 + r_0^2 - 2  r_0\r
\cos\phi)^{-n/2}\ 
\nonumber \\
& = &  1 + a k N(r^2+r_0^2+2 r_0 \r)^{-n/2} F\Bigl(\ha,{n\ov 2},1,{4 r_0 \r\ov
r^2+r_0^2+2 r_0 \r}\Bigr) \ .
\label{haHint}
\ea
The harmonic functions for (\ref{hans55}), (\ref{h44})
reduce in the large $N$ limit to those obtained using (\ref{haHint}).
The result for $n=1$ will be further used in the appendix in connection 
with multi-instanton solutions of 4-dim gravity.
The general harmonic function (\ref{haH}) as well as its derivatives, 
behave as ${a k N\ov r^n}$ for large $r$'s as expected. What is less obvious 
is that very close to the ring (\ref{haHint}) behaves as 
a single-center harmonic smeared out completely along a
transverse direction \cite{sfehi}. 
In other words our multicenter harmonic in $E^{n+2}$
reduces to a single-center harmonic  in $E^{n+1}$.

\section{Branes on a disc}

Consider $N$ static D3-branes distributed, uniformly in the angular direction,
inside a disc of radius $l$ in the  $x_5$-$x_6$ plane.
Their centers are chosen at \cite{sfehi}
\ba
&& \vec x_{ij} = (0,0,0,0,r_{0j} \cos\phi_i, r_{0j} \sin\phi_i )\ ,
\nonumber \\
&& \phi_i = {2\pi i\ov N}\ ,\quad r_{0j}= l \left(j/\sqrt{N}\right)^{1/2}\ ,
\quad  i,j =0,1,\cdots,\sqrt{N} -1\ .
\label{vijk}
\ea
Since we are mainly interested in the large-$N$ limit we may take $\sqrt{N}=
{\rm integer}$ without loss of generality.  
Then, the corresponding harmonic function becomes
\ba 
H_4 & =& 1+ 4\pi g_s l_s^4 \sum_{i,j=0}^{\sqrt{N}-1}{1\ov \left( r^2+r_{0j}^2 
-2 \r r_{0j} \cos(\phi_i-\psi)\right)^2} 
\nonumber \\
&\approx & 1+ {8\pi g_s l_s^4 N\ov \sqrt{(r^2+l^2)^2-4 l^2 \r^2}
\left(r^2-l^2+\sqrt{(r^2+l^2)^2-4 l^2 \r^2}\right)}\ ,
\nonumber\\
&&r^2=x_1^2+\cdots +x_6^2\ ,\qq \r^2=x_5^2+x_6^2\ ,
\label{abcd}
\ea
where the second line is an approximation, valid for large $N$.
The harmonic function $H_4$ becomes singular in the $x_5$-$x_6$
plane inside a disc of radius $r=\r=l$. 
The appropriate supergravity solution 
corresponds to the extremal limit of a rotating D3-brane solution of  
type-IIB supergravity that was found in \cite{russo} (using 
earlier work in \cite{cvsen}).
The radius $l$ of the ring becomes the angular momentum parameter
of the rotating solution.
We emphasize that a priori it is not 
obvious that a non-extremal version 
of the BPS solution with harmonic (\ref{abcd}) should exist at all,
since non-BPS branes exert forces against one another.
In our case the  
gravitational attraction, which is no longer balanced by just the
R--R repulsion, is now balanced by introducing angular momentum.
It is tempting to attribute this balance to some kind of centrifugal force
due to the rotation, but one should remember that this is intrinsic 
and not orbital angular momentum.\footnote{I thank D. Youm for a discussion
on this point.}
It would be interesting to extend the discussion of this section for the 
D3 brane solution with three rotating parameters \cite{KLTri}.

\section{D5's and NS5's on a circle}
Now we specialize to the case of $k N$ D5-branes 
on a circle of radius $r_0$ in the decoupling limit
\ba
&&u_i ={x_i\ov l_s^2}={\rm fixed}\ , \qq U^2 = u_1^2+ u_2^2 + u_3^2 + u_4^2\ ,
\qq u^2= u_3^2+ u_4^2\ ,
\nonumber \\
&&  
g_{\rm YM}^2= g_s l_s^2= {\rm fixed}\ , \qq U_0={r_0\ov l_s^2}={\rm fixed}\ ,
\qq l_s\to 0\ .
\label{lllii}
\ea
We may take 
$r_0\sim l_s/g_s^{1/2}$ or $U_0\sim 1/g_{\rm YM}$ since
the coupling constant $g_{\rm YM}$ is the only scale in the classical theory.
In this limit the appropriate supergravity solution
is (we omit the R--R 3-form magnetic field strength)  
\ba
{1\ov l_s^2} ds^2 & =& {V\ov \sqrt{g_{\rm YM}^2 Nk}} \  ds^2(E^{1,5})
+ {\sqrt{g_{\rm YM}^2 Nk}\ov V}\ du_idu_i \ ,
\nonumber \\
e^{\Phi} &=& {g_{\rm YM} V\ov \sqrt{Nk}}\ ,
\label{d55}
\ea
where $V$ is a function of $U$ and $u$ defined as
\be
V(U,u)= \left((U^2+U_0^2)^2-4 U_0^2 u^2\right)^{1/4} 
\L_N^{-1/2}(x,\psi)\ ,
\ee
and $e^x$ is given by the corresponding expression in (\ref{hans55}) with
$r$, $\r$ and $r_0$ replaced by $U$, $u$ and $U_0$ respectively.
The S-dual description of (\ref{d55}) is in terms 
of $kN$ NS5-branes with a supergravity
description that uses the same harmonic function.
The corresponding metric, antisymmetric tensor field strength and dilaton are
\ba
{1\ov l_s^2} ds^2 & =& ds^2(E^{1,5}) + Nk V^{-2} \ du_i du_i\ ,
\nonumber \\
{1\ov l_s^2} H_{ijk}& =& Nk  \e_{ijkl} \del_l V^{-2}\ ,
\label{ns55}\\
e^{\Phi}& =&  {\sqrt{Nk}\ov g_{\rm YM} V} \ ,
\nonumber
\ea
and represent an axionic instanton \cite{cahastro}.
Note that for energy regimes $g_{\rm YM} U -1\geq {1\ov N}$,
the factor $\L_N$ in the expression for $V(U,u)$
can be ignored and be set to 1.
Using the above, and assuming that $N\gg 1$ and that $U_0\sim 1/g_{\rm YM}$,
we may determine the energy ranges where 
the supergravity description or the ``perturbative'' 6-dim SYM theory one, 
are valid. The discussion parallels the one performed 
in \cite{itzaki} for the 
one-center solution and can be found in \cite{sfehi}.


The geometrical interpretation of the solution (\ref{ns55})
is that of a semi-wormhole with a fat throat and 
$S^3$-radius $\sqrt{Nk} l_s$ \cite{sfe1,sfehi}.
However, as we get closer to any one of the centers, the solution 
tends to represent 
the throat of a wormhole with $S^3$-radius $\sqrt{k} l_s$. Hence, we 
think of (\ref{ns55}) as a superposition of ``microscopic''
semi-wormholes distributed around a circle.
This is to be 
contrasted with the zero size throat of the usual $SU(2)\times U(1)$ 
semi-wormhole. 

\subsection{Exact conformal field theory description}
In the case of NS5-branes, 
we may find an exact CFT description for the background (\ref{ns55})
in two limiting cases. For $N=1$, corresponding to $k$ NS5-branes at a single
point, it is known that the exact description is in terms of the $SU(2)_k
\times U(1)_Q$ WZW model, where $Q=\sqrt{2\ov k+2}$ is the background charge 
associated with the $U(1)$ factor \cite{sl2u1,cahastro}.
An exact CFT description is also possible when $N\gg 1$ \cite{sfehi}.
After a change of variables and a T-duality transformation with 
respect to the vector field $\del/{\del \tau}$ we obtain a 
solution of type IIA supergravity, with the same six 
flat directions
as in (\ref{ns55}), and a non-trivial transverse part given by \cite{sfehi}
\ba
&& {1\ov N k} ds^2_{\perp} = \L_N \bl(d\r^2 + \coth^2\r d\psi^2\br) 
+ \L^{-1}_N (\coth^2\r + \tan^2\th) d\tau^2 
\nonumber \\
&&\qq 
+ \ \L_N \left(1+(1+ \coth^2\r \cot^2\th){\sin^2(N \psi)\ov\sinh^2(N x)}\right)
d\th^2 +\  2 \coth^2 \r \ d\tau d\psi 
\nonumber \\
&& \qq +\ 2 \cot\th {\sin(N\psi)\ov \sinh(N x)}
\bl( \L_N \coth^2\r d\psi + (\coth^2\r + \tan^2\th)  d\tau \br) \ d\th \ ,
\label{dfg}\\
&& e^{-2 \Phi} = { g_{\rm YM}^2 U_0^2\ov Nk}\ \cos^2\th \sinh^2 \r\ ,
\nonumber
\ea
and zero antisymmetric tensor.
In the limit $N\gg 1$, we obtain 
\ba 
&& {1\ov Nk }ds^2_{\perp} 
= d\th^2 + \tan^2{\th}\ d\vphi^2 + d\r^2 + \coth^2 \r\ d\om^2\ ,
 \nonumber \\
&& e^{-2\Phi}={ g_{\rm YM}^2 U_0^2\ov Nk}\ \cos^2\th \sinh^2 \r\ ,
\label{ccoos}
\ea
where $\om = \tau+\psi$ and $\vphi=\tau$. This is 
the background corresponding to the exact CFT
$SU(2)_{kN}/U(1) \times SL(2,\IR)_{kN+4}/U(1)$.
In the opposite extreme case of $N=1$, it can be shown 
that (\ref{dfg}) reduces to 
the background for the exact CFT $SU(2)_k/U(1)
\times U(1) \times U(1)_Q$.
This is no surprise, since the latter background and the one
for $SU(2)_k\times U(1)_Q$ are T-duality related \cite{RSS}.

For the supersymmetric properties 
of the solutions
(\ref{ns55})-(\ref{ccoos}) as well as for their relation to pure 
gravity backgrounds and 
non-extremal black holes arising upon toroidal compactification we refer 
the reader to \cite{sfehi}.

\subsubsection{CFT for non-extremal NS5 branes:}
For completeness we include the discussion of the CFT related to 
a certain limit of the supergravity solution representing
$N$ non-extremal one-center NS5 branes \cite{nonens,nonens5}. 
The corresponding metric, antisymmetric tensor field 
strength and dilaton are \cite{hostro}
\ba
ds^2 &=& ds^2(E^{5}) -f dt^2 +  H \left( f^{-1} dr^2 + r^2 d\Om_3^2 \right)\ ,
\nonumber \\
H_{ijk} &=& \e_{ijkl} \del_l H^\prime  \ ,
\label{ejh}\\
e^{2\Phi} &=& g_s^2 H\ ,
\nonumber
\ea
where 
\ba
&& H = 1 + {\m^2 \sinh^2 \a\ov r^2} \ ,\quad 
 H^\prime = 1 + {\m^2 \sinh\a \cosh \a \ov r^2} \ ,
\quad f=1-{\m^2\ov r^2}\ ,
\nonumber \\
&& \sinh^2\a = \left((\a' N/\m^2)^2 +1/4\right)^{1/2} - \ha \ ,
\label{jhsd}
\ea
where $\m$ is the non-extremality parameter. 
Consider first the change of variables $r=\m \cosh \r$ and then the limit 
$\m,g_s\to 0$ in such a way that ${\m \ov g_s l_s}\equiv r_0$  is held 
fixed.
Then (\ref{ejh}) becomes
\ba
ds^2 &=& ds^2(E^{5}) - \tanh^2 \r\ dt^2 + \a' N d\r^2 + \a' N d\Om_3^2 \ ,
\nonumber \\
H &=& -2 \a' N \e_3 \ ,
\label{ejh1}\\
e^{\Phi} & =& {\sqrt{N}\ov r_0 \cosh \r}\ ,
\nonumber
\ea
where $\e_3$ is the volume form of the unit 3-sphere. 
The background (\ref{ejh1}) corresponds to the $SL(2,\IR)/SO(1,1) \times 
SU(2)$ exact CFT, with the first factor representing the famous 2-dim 
black hole solution \cite{witbla}. 
Note also that, choosing $N$ and $r_0$ in such a way that the
relation $1\ll N \ll r_0^2$ is satisfied, we suppress perturbative as
well as string loop corrections. 

\section{Final comments}

In \cite{sfehi} NS5-branes of type II and heterotic string 
theory whose non-trivial 4-dim 
part is described by the non-Abelian dual 
of 4-dim hyper-K${\rm \ddot a}$hler metrics with $SO(3)$ isometry,
were constructed. In the case corresponding to the non-Abelian dual of 4-dim
flat space the discrete distribution of the brane centers was found.
It will be interesting to extend this analysis to more general 
non-Abelian duals of 4-dim hyper-K${\rm \ddot a}$hler metrics.

It will be interesting to compute the heavy 
quark--antiquark potential for the broken 
${\cal N}=4$ SYM in the large-$N$ limit 
using the AdS/CFT correspondence.
The prototype computation was done for unbroken gauge group in \cite{reymal},
and for 
$SU(N)$ broken to $SU(N/2) \times SU(N/2)$ in \cite{wami}.
In the latter case a ``confining'' behaviour, albeit unstable, was found.
Further breaking the symmetry group 
might result in a stabilization of this behaviour.
It will be interesting to repeat these computations using the 
different supergravity solutions we have presented in this note.

\bigskip
\centerline{\bf Acknowledgements }
\noindent
I would like to thank the organizers of the conference in Corfu (Greece) 
for the invitation to present this and
related work.

\appendix
\section{Comments on gravitational multi-instantons}
\setcounter{equation}{0}
\renewcommand{\theequation}{\thesection.\arabic{equation}}

Consider 4-dim self-dual metrics with a translational Killing vector field
$\del/{\del \tau}$. The general form of the metric is \cite{gravm}
\ba
ds^2 &= &V (d\tau + \om _i dx_i)^2 + V^{-1} dx_i dx_i \ ,
\nonumber \\
\del_i V^{-1} & =& \e_{ijk} \del_j \om_k\ ,\qq i=1,2,3\ .
\label{jhw}
\ea
Hence $V^{-1}$ is a harmonic function of the form 
\be 
V^{-1}= \e + \sum_{i=1}^N {m\ov |\vec x - \vec x_i|}\ .
\label{kas}
\ee
The analogous anti-self-dual metrics can be obtained by the sign change 
$\tau\to -\tau$, so that they will not be considered any further.
With the above choice the singularities at $\vec x= \vec x_i$ are removable 
NUT singularities provided that the variable $\tau$ has period $4 \pi m$.
Hence, it follows that if 
the constant $\e\neq 0$ (in which case it can be normalized to 1)
the space is asymptotically locally flat (ALF). However, if $\e=0$, then 
it is asymptotically locally Euclidean (ALE) with boundary at infinity
$S^3/R_N$ with $R_N$ being a discrete subgroup of $SO(4)$.
Here we are interested in a distribution of a large number $N$ of 
NUT singularities on a circle of radius $r_0$
in a similar fashion as in section 2. Then, except for distances
$r \simeq r_0 + {\cal O}({1\ov N})$ and smaller,
the harmonic function $V^{-1}$ is essentially given 
by (\ref{haHint}) for $n=1$. Using well known formulae we write
\def\rl{r_{{}_<}}
\def\rr{r_{{}_>}}
\ba 
&&V^{-1} \ \approx \
{2 m N\ov \pi} {K(k)\ov (r^2 + 2 r_0 r \sin \th + r_0^2)^{1/2}}
\ =\ mN\sum_{l=0}^\infty {\rl^l\ov \rr^{2 l+1}} P_{2 l}(0) P_{2 l}(\cos\th) \ ,
\nonumber \\
&& k\ =\ 2 \sqrt{ r_0 r\sin\th \ov r^2 + 2 r_0 r \sin \th + r_0^2}\ ,\qq
P_{2l} (0)\ = \ {(-1)^l (2 l)! \ov 4^l (l!)^2}\ ,
\label{ksd}
\ea
where $K(k)$ is the complete elliptic integral of the first kind, $P_l$ is the
usual Legendre polynomial and 
$\rl$ ($\rr$) denotes the smaller (larger) of $r$, $r_0$.
Actually, the expression above 
is a classic result of electrostatics as it represents
the potential due to a uniformly charged ring of radius $r_0$ and total
charge $mN$.
In order to compute the $\om_i$'s appearing in (\ref{jhw}) it is convenient 
to use spherical coordinates $r,\th$ and $\phi$ in which $\om_r =\om_\th=0$. 
The remaining non-vanishing component
$\om_\phi(r,\th)$ is determined by solving
the differential eqs. 
\be 
\del_\th \om_\phi = r^2 \sin \th \del_r V^{-1}\ ,\qq
\del_r \om_\phi = -\sin\th \del_\th V^{-1}\ .
\label{kaH}
\ee
The most convenient way we found to present the solution is as an infinite 
series in terms of Legendre polynomials
\be
\om_\phi = m N \sum_{l=0}^\infty C_l(r) P_{2l+1}(\cos\th)\ ,
\label{saqe}
\ee
where
\ba 
C_l & = & {2 l+1\ov 4 l +1} P_{2l}(0) \left(r_0\ov r\right)^{2l}
-\ {2 l+3\ov 4l +5} P_{2l+2}(0) \left(r_0\ov r\right)^{2l+2}  ,\qq {\rm if}\ 
r> r_0\ ,
\nonumber \\
C_l & = &- {2 l\ov 4 l +1} P_{2l}(0) \left(r\ov r_0\right)^{2l+1} 
+\ {2 l+2\ov 4l +5} P_{2l+2}(0) \left(r\ov r_0\right)^{2l+3}  ,\qq {\rm if}\
r< r_0\ ,
\label{kje}
\ea
Computing the infinite sum explicitly in (\ref{saqe}) seems a difficult 
task to perform. Note that in the limit $r_0\to 0$ only the coefficient 
$C_0=1$ survives. Then $\om_\phi = m N\cos \th$, which is as expected 
for a one-center solution of NUT charge $mN$.


\end{document}